# Using SOA with Web Services for effective Integration of Hospital Information Systems via an Enterprise Service Bus

**Quist-Aphetsi Kester, MIEEE**
**Faculty of Informatics, Ghana Technology University College,**
**PMB 100, Accra-North, Tesano, Accra, Ghana**

**Abstract**

Hospitals are distributed across geographical areas and it is important for all hospitals to share information as well as integrate their systems for effective researching and health delivery. Health personals and institutions in need of information from hospitals with respect to geographical areas can easily do researches on patients, treatments, disease outbreaks, and effects of drugs.

This research work is aimed at integrating of database systems of hospital across geographical areas via a service bus. A centralized service bus was used to facilitate interoperability of applications across platforms and enhance communication within the hospital infrastructure as well as creating enabling environment for new layer of abstractions to be added without modification of the entire system. Concept of Service Oriented Architecture with web services was used for rapid integration solution in solving the challenges faced during integration of multiple incompatible applications.

## 1. Introduction

To achieve the goal of the global national e-health strategy that intends to provide an interoperable, standardized and secure platform for all involved partners in supporting healthcare services, one needs to focus on interoperability and integration of all distributed enterprise health information management systems that exists across geographical areas. This should cover hospitals, health research institutions etc. Having access to the right information in real-time is among the most efficient ways of decision making in today's world. GIS as part of a information systems can be used in generating geographical reports for effective analysis and decision making. [1][2]

Service Oriented Architecture (SOA) has radically changed the application integration landscape. SOA can be considered as a business-centric approach for enabling integration. Visibility, interaction and effect are the key concepts in any SOA implementation [3]. Visibility refers to the capacity of those with a need to see those with a capacity to service the needs. This is accomplished by publishing the service in a registry.

As corporate dependence on technology has grown more complex and far reaching, the need for a method of integrating disparate applications into a unified set of business processes has emerged as a priority. Users and business managers are demanding that seamless bridges be built to join them. In effect, they are demanding that ways be found to bind these applications into a single, unified enterprise application. The development of Enterprise Application Integration (EAI), which allows many of the stovepipe applications that exist today to share both processes and data, allows us to finally answer this demand. [4]

Service Oriented architecture (SOA) has gained popularity in recent years due to its enabling functionality or services to upgrade and extend existing software applications. SOA is an architectural approach to build and deploy software applications that is interoperable by design.SOA has grown as companies endeavor to leverage their existing client base and to integrate their acquired software with their clients' existing ERP system and also it makes software connectivity capabilities very easy. Unlike EAI no middleware is needed as adoptions of standards enable services to interact directly. It also enhances reusability capacity of software, resulting longer life of existing assets. A successful SOA implementation makes it easier to customize and upgrade existing applications thereby reducing total cost of ownership. [5] The adoption of service oriented Architecture and web services provide a rapid solution to solving integration problems faced by organizations [6] [7] [8] [9].

The paper has the following structure: section 2 consist of related works, section 3 gives information on the methodology, section 4 discusses the approach used for the integration, 5 talks about implementation as well as results and section concluded the paper.

## 2. Related Works

In this development paradigm, functionality is exposed as services thereby enabling service requesters and providers interact through messages. Services are built to be autonomous but can also be combined to form even larger services and applications. Service orientation provides guidelines and principles that govern the creation, implementation and management of services.

Primarily, services are implemented as Web Services (WS) which are defined by the W3C as "software systems designed to support interoperable machine-to-machine interaction over a network" [10].It has an interface described in a machine-processable format. Other systems interact with the Web service in a manner prescribed by its description using SOAP-messages, typically conveyed using HTTP with an XML serialization in conjunction with other Web-related standards [10].Extensible Markup Language (XML) has emerged as a powerful self-describing language to enable businesses to share information and conduct transactions on the Internet. The emergence of XML as a standard, to a large extent, has driven the evolution of application integration technologies. [1]

Inherently, a service is a software component that contains a collection of related software functionalities reusable for different purposes [11].It delivers such operations as data storage, data processing, mathematical and scientific computations, and networking. It is governed by a producer-consumer model in which a service is delivered by a service provider known as the producer which owns the facilities for hosting, running, and maintaining the service, and the client known as the consumer which connects and uses service functionalities.

Much of the research regarding SOA tackles more granular technical issues of development and implementation of Web services, which may be a result of the aforesaid misconceptions [12]. Few papers e.g., [13],[14], deal with the much larger problem of defining what SOA means to the organization and how this definition should then provide the guidance for the development of components to meet business information needs[15]. The IT adoption literature targeting a methodology for development states that there are five categories of factors influencing the decision to adopt SOA (i.e., environmental, organizational, individual, technology, and task characteristics [16]. These same factors should be addressed by the methodology for implementing SOA projects [17]. We now discuss two SOA methodologies that attempt to embody some or all of these factors.

Teti (2006), an industry analyst, provides a methodology, which entails creating a vision, construction, and execution. He suggests that this model is applicable to many projects, but specifically addresses SOA. The vision creation is driven by a number of inter- and intra-organizational issues that define tasks important to the individuals and the firm (i.e., the constituency); the construction addresses the technology required to accomplish the tasks; and execution seeks to ensure that SOA will facilitate information exchange in the environment.

Bell (2008) provides a SOA methodology that takes a more technical approach. It professes that all software can be considered as services that are designed based on the informational tasks of the organization, configured for transmission in the working environments, constructed with available technologies, and deployed for use by individuals. The methodology represents a conceptual structure that brings together distributed services based on the functionality [18].

## 3. Methodology

SOA with web services guides all aspects of creating and using software services throughout the software system lifecycle Thus, from their conception to their retirement. And also in defining and conditioning the IT infrastructure that allows different applications to exchange data and participate in business processes regardless of the operating systems or programming languages underlying those applications. [6] [19]

This research work is aimed at integrating of hospital information systems of hospitals via a service bus. A centralized service bus was used to facilitate interoperability of applications across platforms and enhance communication within the hospital infrastructure as well as creating enabling environment for new layer of abstractions to be added without modification of the entire system. The entire integration process was based on the concepts of SOA with web services. The form of approach used was to enable other health services within and outside hospitals to communicate and exchange data.

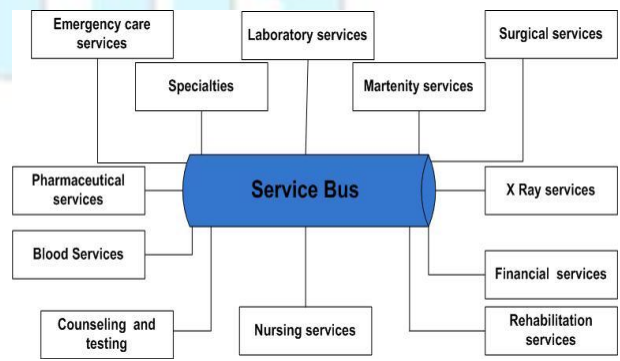

Fig. 1 Service integration model for hospitals

In the above figure, the services of the hospitals were connected via a central service bus to enable information exchange and effective data communication between software applications. This also enabled interoperability for heterogeneous applications where multiple systems can share information. This further makes data migration and usage easy from one system to another.

## 4. The Integration Approach

A multi-tier architecture was employed in devising the solution. The architecture consists of the database layer, business logic layer, middleware/ service bus, and presentation layer. The database and business logic layer reside at the service providers' end. The database information would be made available to requestors through the data access layer, while the business logic layer is where the business rules that determine how data is created, stored and accessed are implemented. The database technologies and business logic implementations vary across service providers. As a result of the service-oriented architecture being implemented, external access to these databases is provided via services. The services are implemented using web service technology. The choice of web services technology is due to its support for open technologies and protocols such as HTTP (hypertext transfer protocol), XML (extensible markup language), and, WSDL (web service definition language) among others. These open technologies provide a desired level of interoperability and easy means of access. The web services provide functionality to access the databases while implementing the business logic at the backend. The individual services are registered in a repository from where they can be accessed. Users can access services directly in repositories; in this case, access will be through a middleware. An enterprise service bus is the middleware between service requestors and the services providers. Since the requestors need to access various services to get comprehensive information from the hospitals, the ESB performs the task of determining what service to access and the order they are accessed. In other words, the ESB performs service orchestration. Service consumers cannot access the ESB directly so a further interface is required. Service consumers request services today via an array of devices ranging from desktop computers and laptops to mobile devices such as smartphones and tablets. A universal mode of access is required for all these devices hence the choice of a web interface. Most modern devices contain a web browser that can be used to access web content irrespective of operating platforms. As a result, a web server is required to provide a presentation layer in form of web pages to the consumers via which they can access the services and view results.

This choice of architecture is effective because it provides access to a wide range of consumers and devices as a result of the ubiquitous technologies employed in the presentation layer while also providing a high level of separation of concern. However, the consumers are loosely coupled to the individual services. This provides flexibility because the service providers can switch database technologies or business logic implementation without affecting the consumer as long as the exposed services conform to a predefined contract that the consumer is accustomed to. Implementation of services by providers is also relatively cheap as services can be built on existing infrastructures without restructuring the whole system.

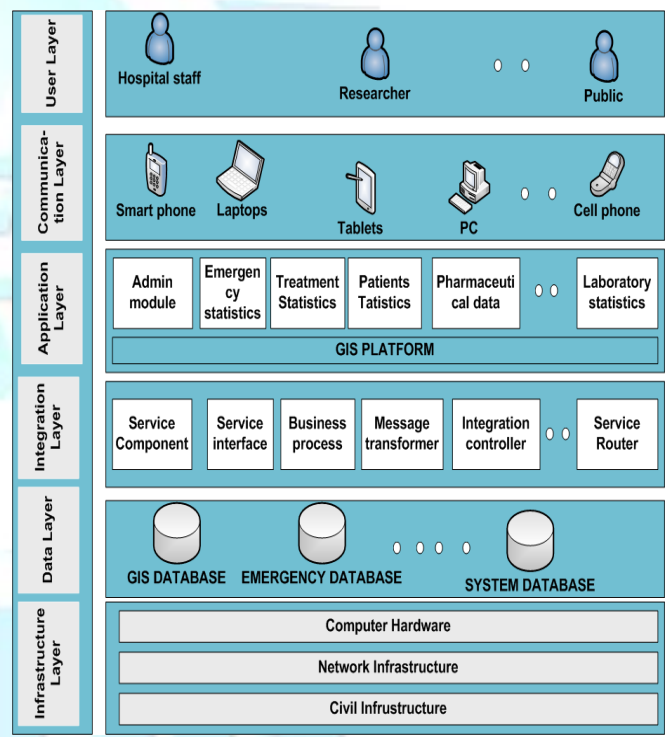

Fig. 2 The layer architecture for hospitals

From figure 2, the Infrastructure Layer consists of the computer hardware (Servers, desktops, peripheral devices, UPS, etc.); civil infrastructure designed for (server rooms, control centers, etc.) and network infrastructure (Structured cabling, switchers, routers, fiber optic channels, etc). The Data Layer is composed by required databases like a GIS database (reference maps, road maps, high resolution imaginary); A database to store logs and records of past and ongoing records; Other related databases and a system database to manage records etc. the integration layer exposes functions to both external and inter users. The Application Layer integrates functions into modules which is made available to communication carriers define in the communication layer. Users Layer represent users of the system.

The Usage of the Integration Layer for hospitals as shown in Figure 3 provides a level of indirection between the consumer of functionality and its provider. A service consumer interacts with the service provider through the Integration Layer. Therefore, each service interface is only exposed via the Integration Layer (e.g., Enterprise Service Bus), never directly and point-to-point integration is done at the Integration Layer instead of consumers/requestors doing it themselves. Consumers and providers are decoupled; this decoupling allows integration of disparate systems into new solutions.

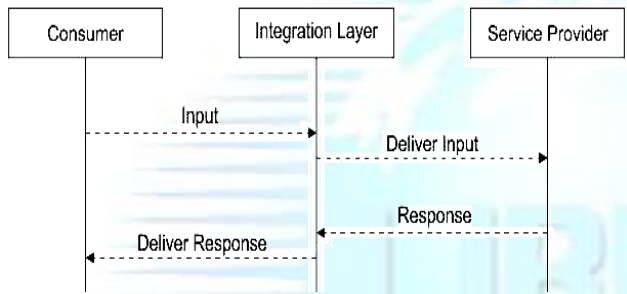

Fig. 3 Usage of the Integration Layer by hospitals

Consumers seeking to use a service are provided with a front end web page through which they can search for information. The search input is posted back to the web server on submission and then the input is transferred to the enterprise service bus via the appropriate adapter. The ESB has a collection of approved web services exposed by the various vendors in its registry. The ESB passes the input parameters to the appropriate services in the registry. These services in turn transfer the input parameter to their remote applications on which their business logic resides. The applications query the databases using the supplied parameter and the results are passed back to the ESB. The results are accumulated and transferred back to the web server where they are formatted and displayed to the end user in a web page.

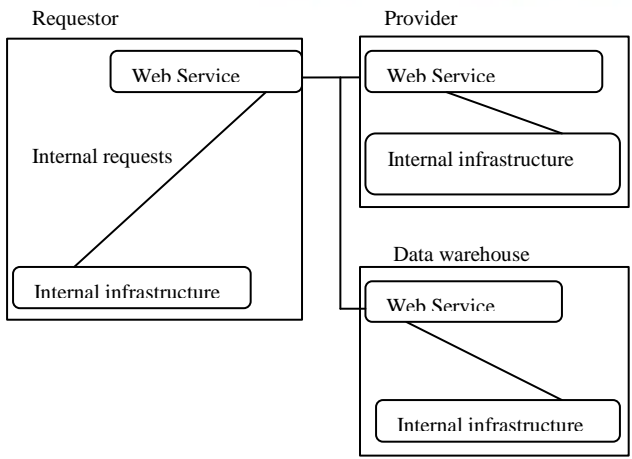

Fig. 4 Web enabled systems for provider and requestor.

In the figure above, languages and protocols were standardized to eliminate the need for many different middleware infrastructures and the interactions were based on protocols redesigned and the internal functionality settings were made available as a service. Service-oriented architecture has standardization as a key policy in implementing web services for easy integration of multiple incompatible applications [9]. Web services provide an entry point for accessing local services and with homogeneous components that reduces the difficulties of integration. Web services were exposed through the interface. Homogeneous components were built to reduce the difficulties of integration. Service descriptions were made richer and more detailed, covering aspects beyond the service interface.

The architecture of the solution proposed in this paper is based on the Open Group SOA Reference Architecture as seen in figure 5 below. The SOA reference architecture provides a baseline that shows the basic layers involved in a typical SOA solution. This diagram shows the different layers of the reference model and how they fit together to provide a standard service oriented solution. The Open Group SOA Ontology provides a taxonomy and ontology for SOA. [20]

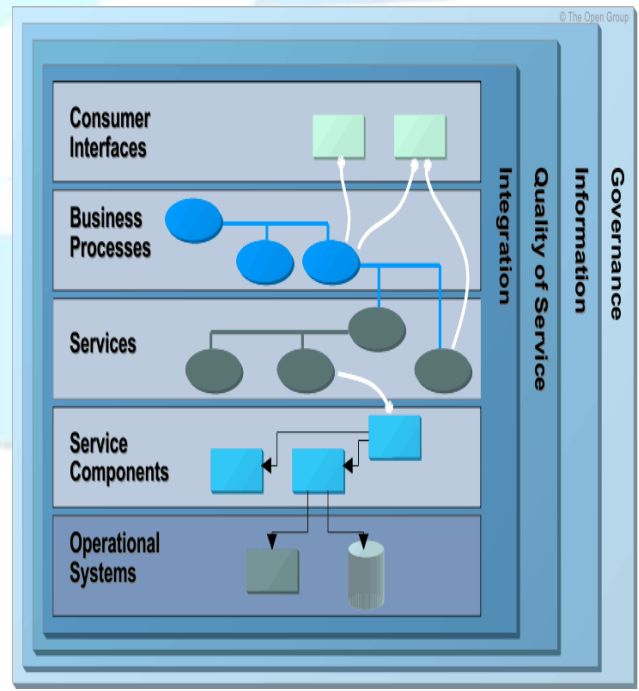

Fig. 5 the Open Group SOA Reference Architecture

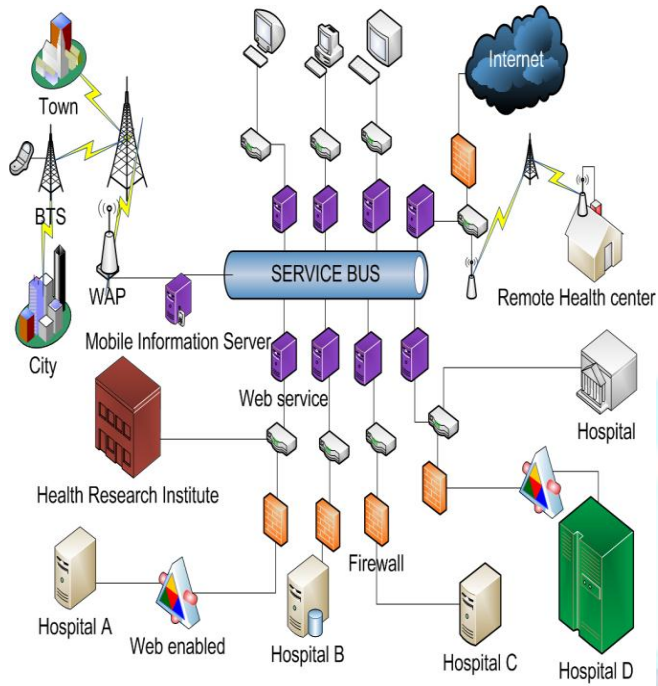

Fig. 6 Integrated Hospital Information Systems via a Service Bus

Figure 6 illustrates how various disparate data sources of hospital information systems are connected via a service bus. Institutions and individuals can access services using Smartphone and other devices.

## 5. Implementation and results

The implementation was done by building the N-tier database application that uses Java servlets and the Java Database Connection (JDBC). The tiers consist of Client Tier or user interface, Middle Tier or business logic and Data Storage Tier. The logical tiers were implemented to correspond to three types of hosts namely: the browser or GUI Application to serve the client, Web Server or Application Server and the Database Server (often an RDBMS or Relational Database). Servlets was used for creating HTML user interface pages. The servlets JavaBeans were responsible for business logic and Java classes responsible for data access. These objects used were JDBC to query the database. The HttpServlet class provides methods, such as doGet and doPost, for handling HTTP-specific services.

The web client consists of two parts: Dynamic web pages containing various types of markup language such as HTML and XML, which are generated by web components running in the web tier and web browser, which renders the pages received from the server. The application clients directly access enterprise beans running in the business tier. The clients interact with Java EE servers, enabling the Java EE platform to interoperate with legacy systems, clients, and non-Java languages.

The middle tier was developed with a Web server running Java servlets which accesses a database and returns an HTML page. Web browsers then communicate with the server by using HTTP. Java servlets requests through methods inherited from the HttpServlet class: doGet(HttpServletRequest, HttpServletResponse) and doPost(HttpServletRequest, HttpServletResponse). The doGet() and doPost() methods have two arguments: an HttpServletRequest object and an HttpServletResponse object. The servlet communicates with the user by sending back an HTML document, a graphics file, or other types of information supported by the web browser. It sends this information by calling the methods of the HttpServletResponse class after the request have been made and the results are been sent.

The general steps taken in setting up the back-end database server are as follows:
- Importing the packages
- Registering the JDBC drivers
- Opening a Connection to the Database
- Creating a Statement Object
- Executing a Query and Returning a Results Set Object
- Processing the Result Set
- Closing the Result Set and Statement Objects
- Closing the Connection

At the end web services were exposed through the interface and the functionality performed by the internal systems and this makes the services discoverable and accessible through the Web in a controlled manner. Homogeneous components were built to reduce the difficulties of integration and standardized. The application integration in environment encompasses three layers: a business process layer, an integration layer, and an application server layer. Each layer, in turn, holds technologies that serve as the application server integration building blocks. The application server layer enables an application integration project to link not only with existing enterprise systems but also with the Web. The application integration platform adds an integration layer on top of application server. To integrate applications at the business logic layer, systems were enabled to consume and provide XML-based Web services.

## 6. Conclusions

With the proposed system, individuals as well as health institutions can now search for information from databases that have been web enabled and their services registered within the service repository which is discoverable via the service bus. Reports on outbreak of diseases can easily be traced from databases from hospitals based on reported cases as well as other statistics can be obtained as well. Researchers can also now use information obtained from hospital databases easily and in real-time. Using SOA with web services makes it easy for heterogeneous database platforms to be integrated and interoperate. Services created can be reused in multiple ways and also new services and applications can be created quickly and easily used with a combination of new and old services.